\documentclass[conference]{IEEEtran}
\IEEEoverridecommandlockouts
\usepackage{cite}
\usepackage{amsmath,amssymb,amsfonts}
\usepackage{algorithmic}
\usepackage{graphicx}
\usepackage{textcomp}
\makeatletter

\newcommand{\Rmnum}[1]{\expandafter\@slowromancap\romannumeral #1@}
\makeatother
\usepackage{multirow, makecell}
\def\BigRoman{\uppercase\expandafter{\romannumeral\number\count 255 }}
\def\Romannumeral{\afterassignment\BigRoman\count255=}
\usepackage{tikz}
\def\BibTeX{{\rm B\kern-.05em{\sc i\kern-.025em b}\kern-.08em
    T\kern-.1667em\lower.7ex\hbox{E}\kern-.125emX}}

\begin{document}

\title{Decoding Continual Muscle Movements Related to Complex Hand Grasping from EEG Signals
\footnote{\thanks{This work was partly supported by Institute of Information \& Communications Technology Planning \& Evaluation (IITP) grant funded by the Korea government (MSIT) (No. 2017-0-00432, Development of Non-Invasive Integrated BCI SW Platform to Control Home Appliances and External Devices by User’s Thought via AR/VR Interface; No. 2017-0-00451, Development of BCI based Brain and Cognitive Computing Technology for Recognizing User’s Intentions using Deep Learning; No. 2019-0-00079, Artificial Intelligence Graduate School Program, Korea University).}
}
}

\author{\IEEEauthorblockN{Jeong-Hyun Cho}
\IEEEauthorblockA{\textit{Dept. Brain and Cognitive Engineering}\\
\textit{Korea University} \\
Seoul, Republic of Korea \\
jh\_cho@korea.ac.kr}\\

\IEEEauthorblockN{Byeong-Hoo Lee}
\IEEEauthorblockA{\textit{Dept. Brain and Cognitive Engineering}\\
\textit{Korea University} \\
Seoul, Republic of Korea \\
bh\_lee@korea.ac.kr}\\

\and

\IEEEauthorblockN{Byoung-Hee Kwon}
\IEEEauthorblockA{\textit{Dept. Brain and Cognitive Engineering}\\
\textit{Korea University} \\
Seoul, Republic of Korea \\
bh\_kwon@korea.ac.kr}\\

\IEEEauthorblockN{Seong-Whan Lee}
\IEEEauthorblockA{\textit{Dept. Artificial Intelligence}\\
\textit{Korea University} \\
Seoul, Republic of Korea \\
sw.lee@korea.ac.kr}
}


\maketitle

\begin{abstract}
Brain-computer interface (BCI) is a practical pathway to interpret users' intentions by decoding motor execution (ME) or motor imagery (MI) from electroencephalogram (EEG) signals. However, developing a BCI system driven by ME or MI is challenging, particularly in the case of containing continual and compound muscles movements. This study analyzes three grasping actions from EEG under both ME and MI paradigms. We also investigate the classification performance in offline and pseudo-online experiments. We propose a novel approach that uses muscle activity pattern (MAP) images for the convolutional neural network (CNN) to improve classification accuracy. We record the EEG and electromyogram (EMG) signals simultaneously and create the MAP images by decoding both signals to estimate specific hand grasping. As a result, we obtained an average classification accuracy of 63.6($\pm$6.7)\% in ME and 45.8($\pm$4.4)\% in MI across all fifteen subjects for four classes. Also, we performed pseudo-online experiments and obtained classification accuracies of 60.5($\pm$8.4)\% in ME and 42.7($\pm$6.8)\% in MI. The proposed method MAP-CNN, shows stable classification performance, even in the pseudo-online experiment. We expect that MAP-CNN could be used in various BCI applications in the future.
\end{abstract}

\begin{small}
\end{small}

\section{Introduction}
{B}{rain-computer} interfaces (BCI) is one of the essential and practical approaches to translating human intentions into external device commands from electroencephalogram (EEG) signals. BCI technology development has been studied mainly to avoid indirect manipulation such as using a keyboard or a joystick but directly interacting with external devices by decoding users' intentions \cite{ECoG, ECoG2, C1, C2}. One of the meaningful developing directions in BCI technology yielded remarkable improvement as assisted-living devices for individuals with motor or sensory impairments due to stroke and amyotrophic lateral sclerosis \cite{jeong2020decoding, jeong2020brain, zhang2019strength, zhang2017hybrid, A1}. Therefore, it is valuable to decode movement's intention accurately through BCI and predict commands for device control that can be used in daily lives \cite{karavas2017hybrid, lafleur2013quadcopter}. The eventual outcome of this interaction using BCI might be the construction of synchronization between humans and machines by decoding intuitive intentions with high accuracy and communicating in real-time \cite{tayeb2020decoding, jeong2020brain, MRCP}.

Meanwhile, hand grasping is associated with more dynamic brain activity than the movements of other extremities because a large area of the human brain's motor cortex is allocated to controlling the hands \cite{channel, highertemporal, osborn2018prosthesis}. In particular, decoding grasp actions within a single arm is more complicated than decoding the conventional motor execution (ME) and motor imagery (MI) related to the movements of other body parts. Due to this significance, many researchers who study BCIs attempted to decoding complex movement intentions in different body parts \cite{tayeb2020decoding, ganzer2020restoring, jeong2020brain, kwon2019subject, karavas2017hybrid}.

Recently, numerous studies have attempted to decode EEG signals related to actual movements or motor intentions using diverse approaches \cite{kim2019subject, karavas2017hybrid, wang2018wearable, stawicki2017novel, cho2020decoding, koizumi2018development, won2017motion, kwon2019subject}; consequently, these advances in BCI technology have inspired our present research. The latest studies on BCIs have been aimed at real-time robotic devices and neuroprosthesis control \cite{highertemporal, jeong2020decoding, jeong2020brain, ganzer2020restoring, tayeb2020decoding}. However, no study has attempted to decode continual muscle movements, especially hand grasping in a single arm based on noninvasive BCIs. Hence, this paper focuses on decoding the continual muscle movements related to various hand grasping (e.g., cylindrical, spherical, lateral, and no-movement) and improving previous achievements of our preliminary study \cite{cho2020novel}.

\begin{figure}[t!]
\centerline{\includegraphics[width=\columnwidth]{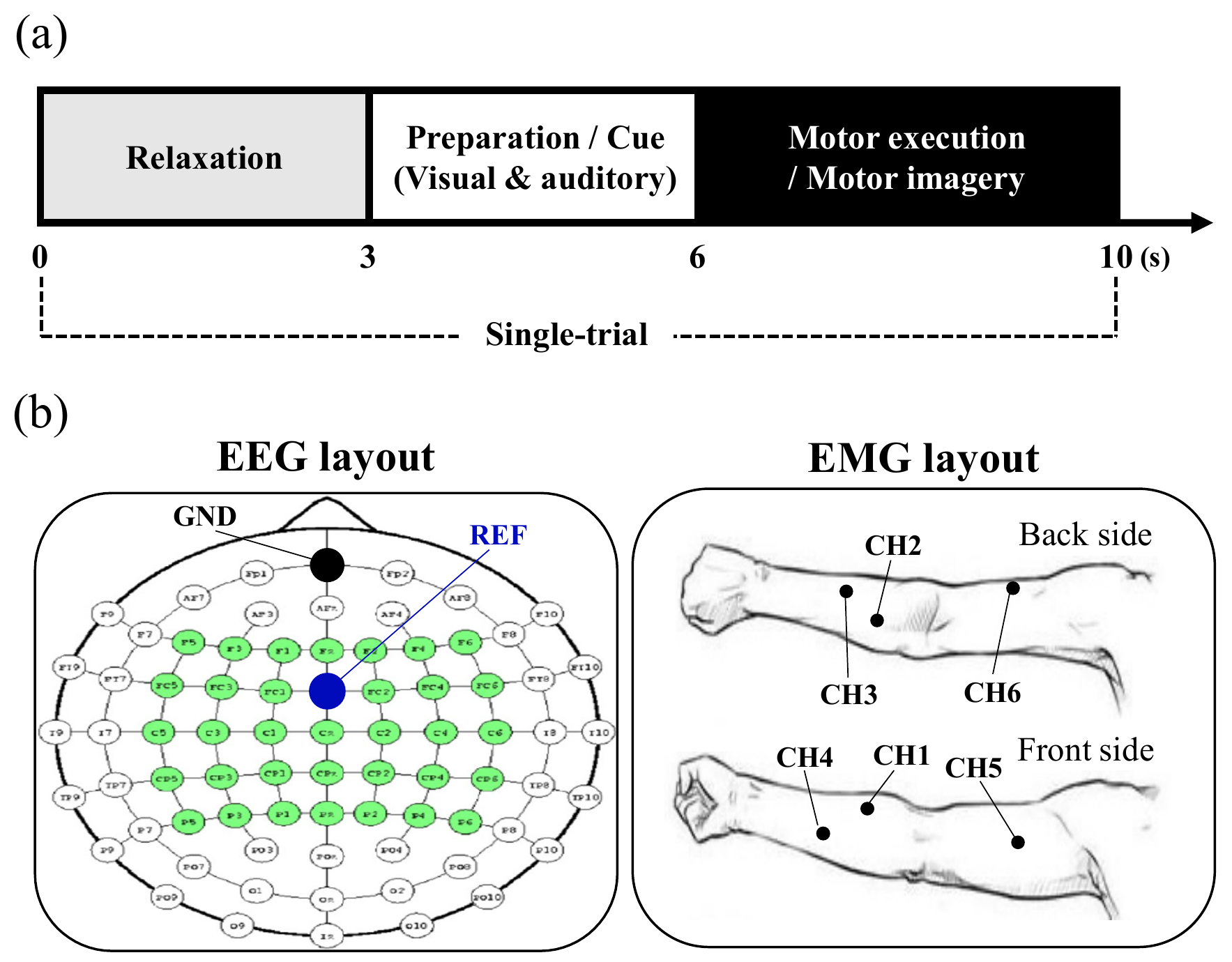}}
\caption{Experimental protocol and electrodes layouts for data acquisition. (a) Experimental protocol; a single trial lasts 10 s, and the subjects perform motor execution and motor imagery during the designated stage for 4 s after the visual cue was given. (b) EEG and EMG layouts.}
\end{figure}
\begin{figure}[t!]
\centerline{\includegraphics[width=\columnwidth]{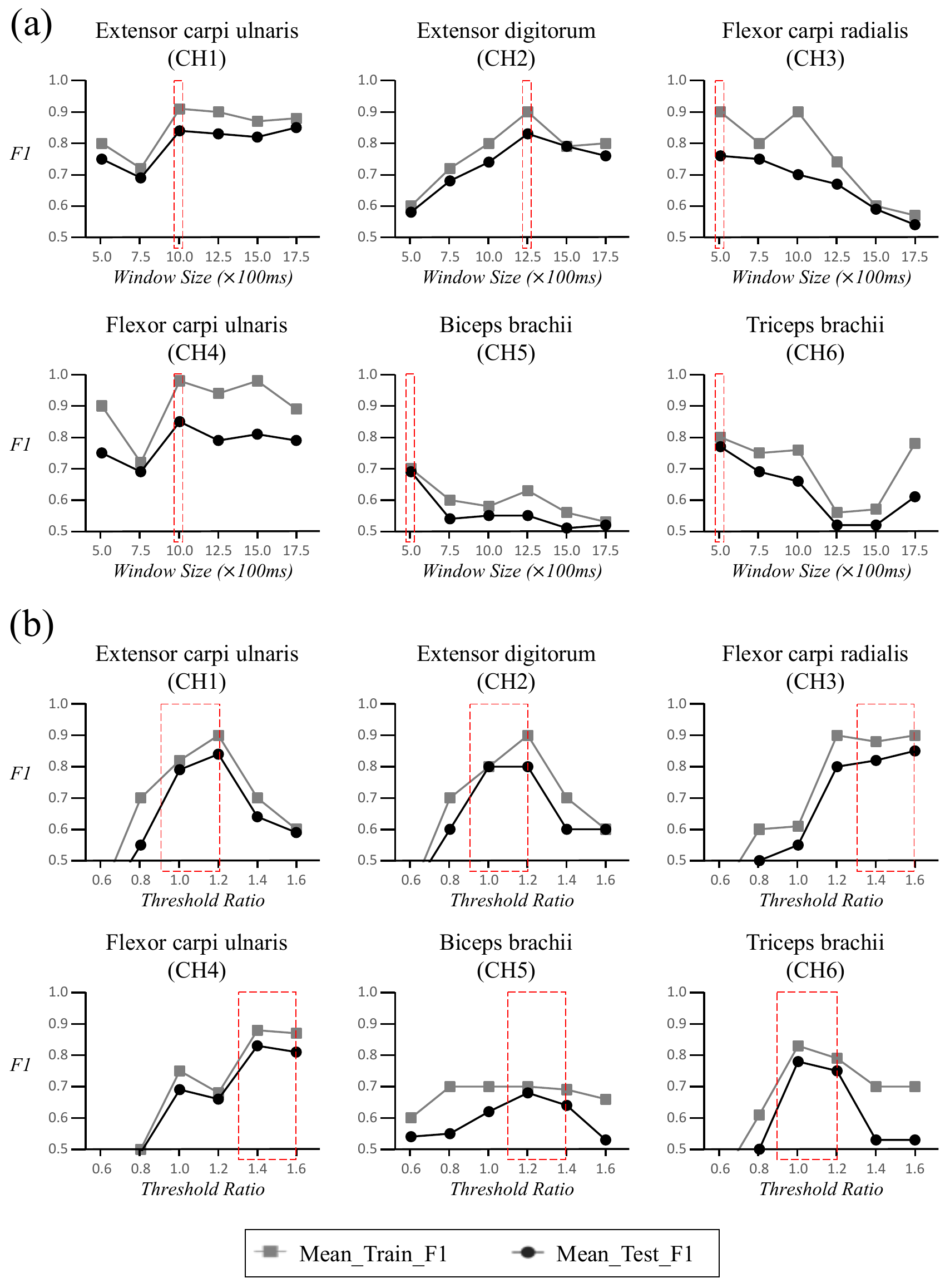}}
\caption{Performance evaluation on detecting muscle activation by EEG decoding. (a) Classification accuracy in F1-score with various window sizes between 500 to 1,750 ms. (b) Evaluation under various threshold ratios, which is the value that determines muscle activation and creates ground-truth labels. We selecting the top four threshold ratios, from 0.6 to 1.6, and applied a step size of 0.1.}
\end{figure}

\section {Materials and Methods}
\subsection{Participants}
Fifteen voluntary subjects (Sub01--Sub15; aged 20--34; all right--handed) who were na{\"i}ve BCI users participated in the experiments. Once they confirmed that they understood the experiment, each subject provided written consent according to the Declaration of Helsinki. The experimental protocols and environments were reviewed and approved by the Institutional Review Board at Korea University [1040548-KU-17-172-A-21].

\subsection{Experimental Setup}
The subjects have followed the protocol presented in Fig. 1(a) during the entire experimental session. The locations of the EEG and EMG electrodes are described in Fig. 1(b). The subjects were asked to perform three intuitive ME and MI about grasping three designated objects (cup, ball, and card): cylindrical, spherical, and lateral grasp, respectively.

\begin{figure*}[t!]
\centerline{\includegraphics[width=\linewidth]{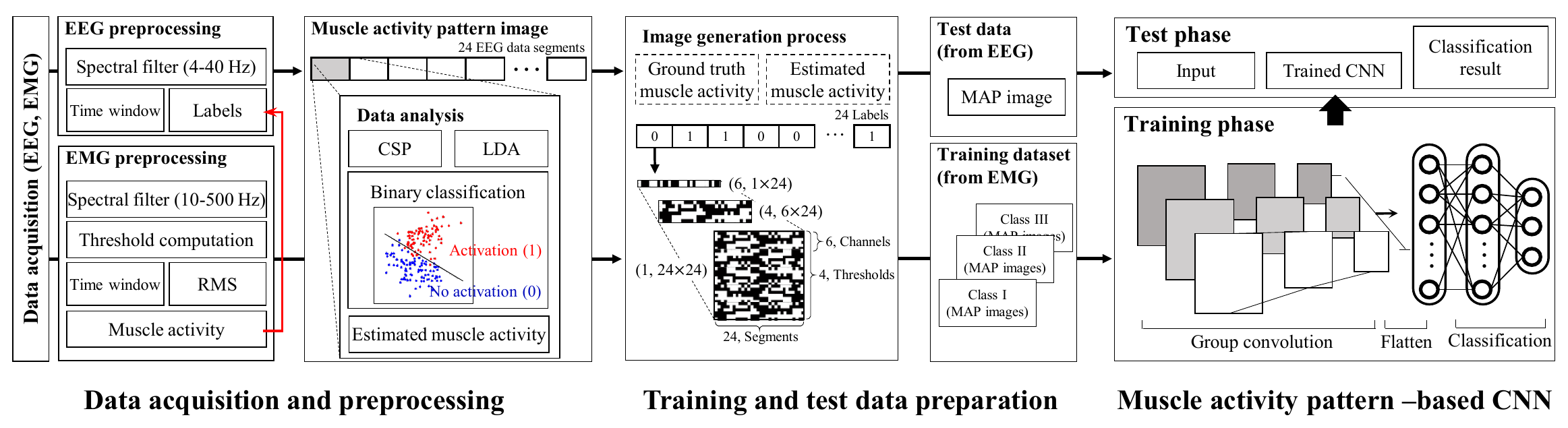}}
\caption{Flowchart of the proposed method, which uses the images of muscle activity pattern created by decoding raw signals. The method consists of the training phase and the test phase.}
\end{figure*}
\begin{figure}[t!]
\centerline{\includegraphics[width=\columnwidth]{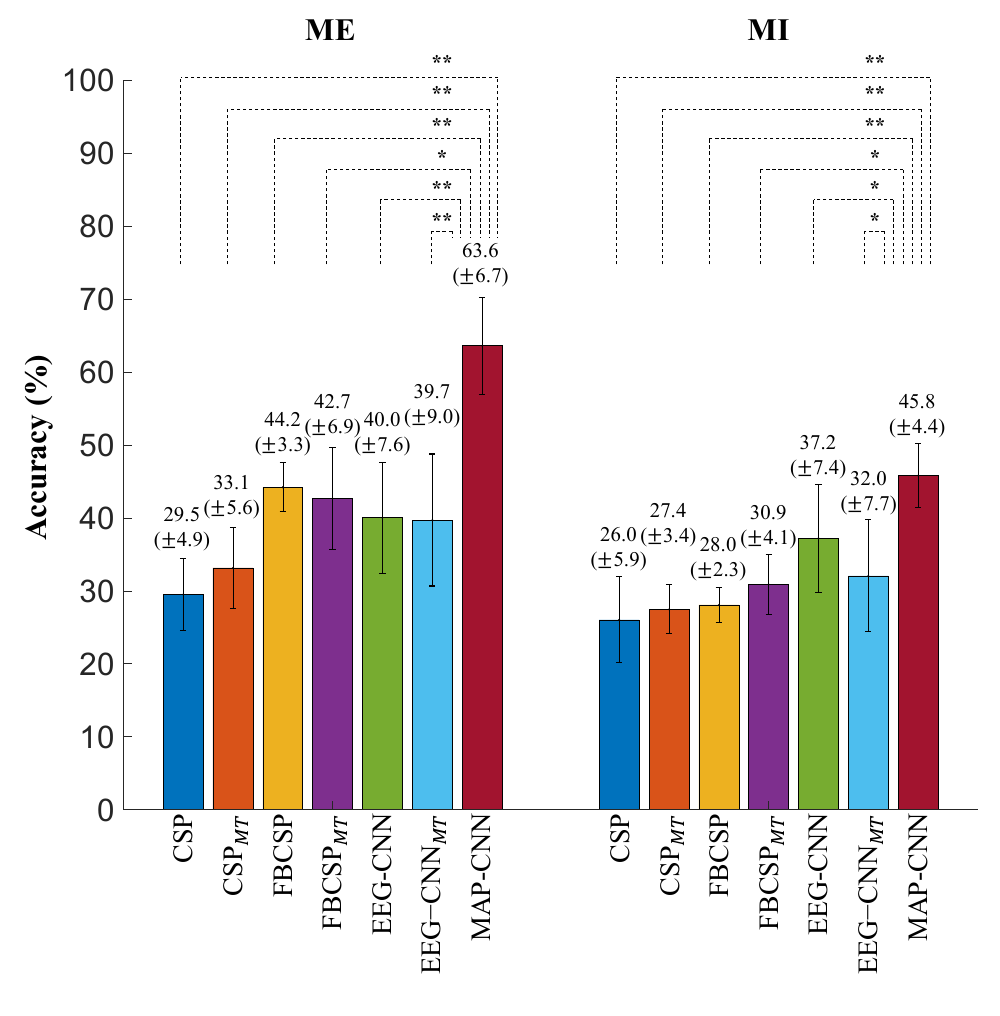}}
\caption{Average classification accuracy results in the ME and MI paradigms, comparison between the proposed and comparable methods in the offline experiment. Note: $^{*}$ and $^{**}$ present a \emph{p}-value from the paired \textit{t}-test ($^{*}$: \emph{p} $<$ 0.05, $^{**}$: \emph{p} $<$ 0.01).}
\end{figure}

\subsection{Data Acquisition}
EEG data were collected at 1,000 Hz using 34 Ag/AgCl electrodes (F1$-$6, Fz, FC1$-$6, C1$-$6, Cz, CP1$-$6, CPz, P1$-$6, and Pz) in the 10/20 international system via BrainAmp (BrainProduct GmbH, Germany) \cite{intuitiveMI, channel, highertemporal, zhang2019strength}. A 60 Hz notch filter was used to remove the power frequency interference. The FCz and FPz were used as the reference and ground electrodes, respectively. The EEG data was bandpass filtered in the range [4--40] Hz \cite{ELU, entropy} using Hamming--windowed zero--phase finite impulse response (FIR) filters with an optimized order (N=50) \cite{spatialfilter}. 

The EMG data of movements in the right arm were recorded from six related muscles, as shown in Fig. 1(b). The details of the related muscles are as follows: extensor carpi ulnaris (CH1), extensor digitorum (CH2), flexor carpi radialis (CH3),  flexor carpi ulnaris (CH4), biceps brachii (CH5), and triceps brachii (CH6).

\begin{figure}[t!]
\centerline{\includegraphics[width=\columnwidth]{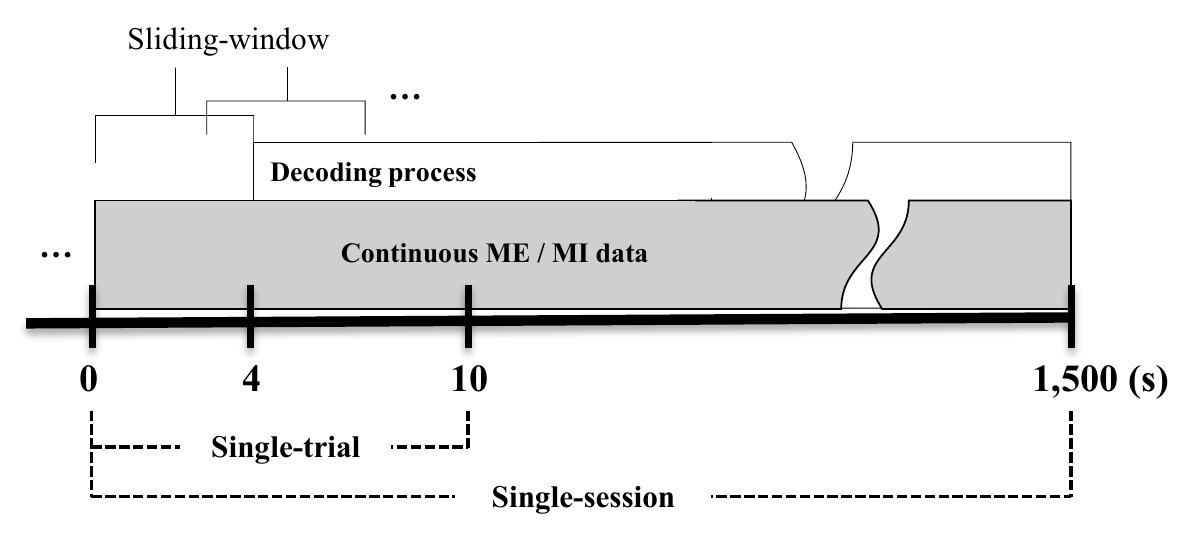}}
\caption{Illustration of the experimental protocol for pseudo-online. Offline data was modified to a continuous EEG data by concatenating all trials. EMG signals were excluded from the original data.}
\end{figure}

\subsection{Data Analysis}
To accurately predict each muscle movement based on EEG data, it is critical to find the optimal sliding window size and threshold for data analysis in advance. Fig. 2(a) presents the optimal EEG data sliding window size for muscle motion decoding, and Fig. 2(b) shows the optimal amplitude threshold of EMG signals to be evaluated as activation or not.

The proposed method consisted of training and test phases. Fig. 3 shows a flowchart of the proposed method that creates input images from two different sources of raw signals. In the training phase, we generated patterns by detecting muscle activation from EMG signals and converting the sum of these classification results into an image that represents the muscle activity pattern (MAP). We estimated the MAP by decoding the EEG signals in the test phase. The convolutional neural network (CNN) \cite{CNN, entropy, ELU} for the final classification was trained with the images collected during the training phase. The trained networks were applied in the test phase to classify the input image into the correct class, representing the estimated MAP taken from EEG signals.

In order to estimate the MAP corresponding to the result of decoding EMG signals in the training phase, the same-sized sliding window was applied to the EEG signals, and we extracted spatial features using the common spatial pattern (CSP) on the segmented data. After the CSP process, the features were used as input data to train the classifier. We used the ground-truth label, which was obtained by detecting muscle activation, to train and test the linear discriminant analysis (LDA) for EEG. Through this process, we estimated the MAP by decoding EEG signals. As a result, we obtained the input MAP images of three groups, corresponding to each grasping class, in order to train the neural network.

\begin{figure*}[t!]
\centerline{\includegraphics[width=\linewidth]{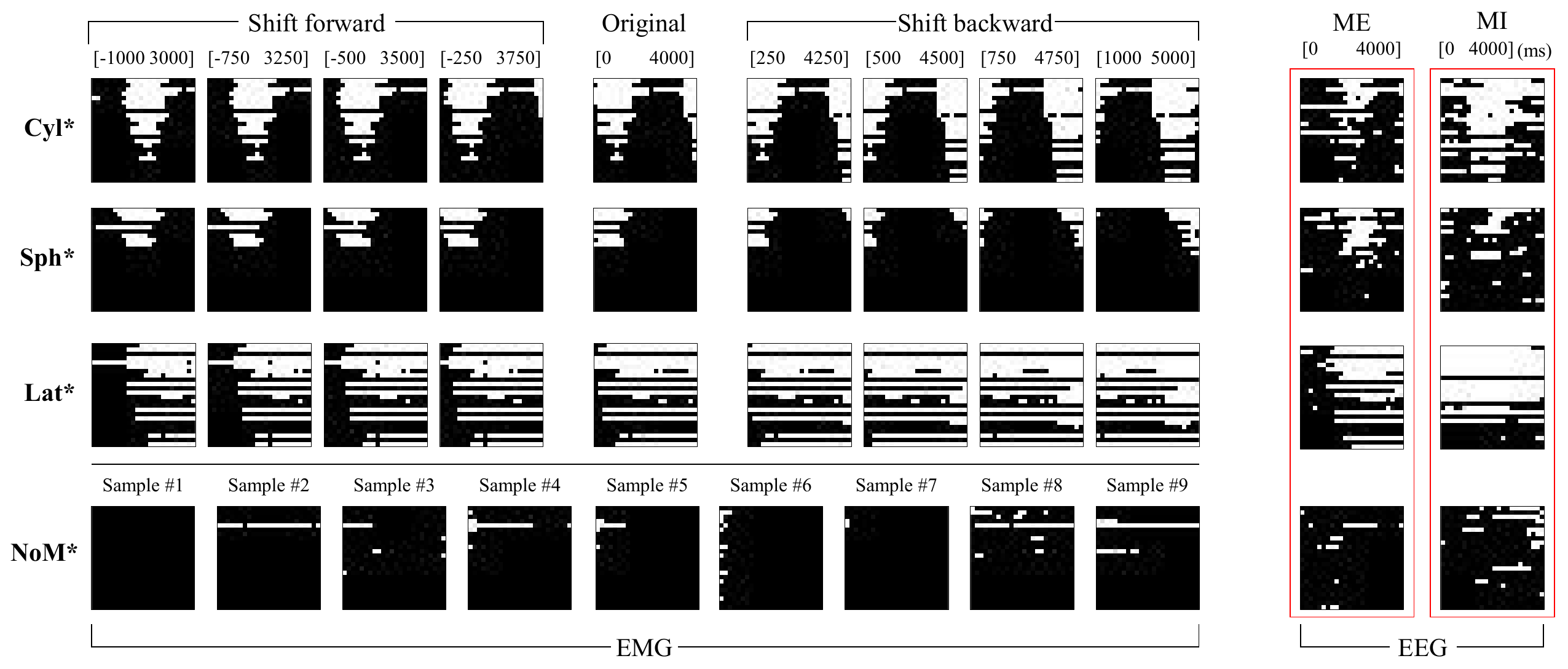}}
\caption{Sample images listed for analysis (from the Sub02). Each category corresponds one of the grasping actions or to no-movement. Time-shift was applied by shifting time interval backward and forward on the original data. The 'EMG' and 'EEG' shown below indicate the source of the signal used to generate the image. *Cyl: cylindrical grasp, *Sph: spherical grasp, *Lat: lateral grasp, and *NoM: no-movement.}
\end{figure*}
\begin{figure}[t!]
\centerline{\includegraphics[width=\columnwidth]{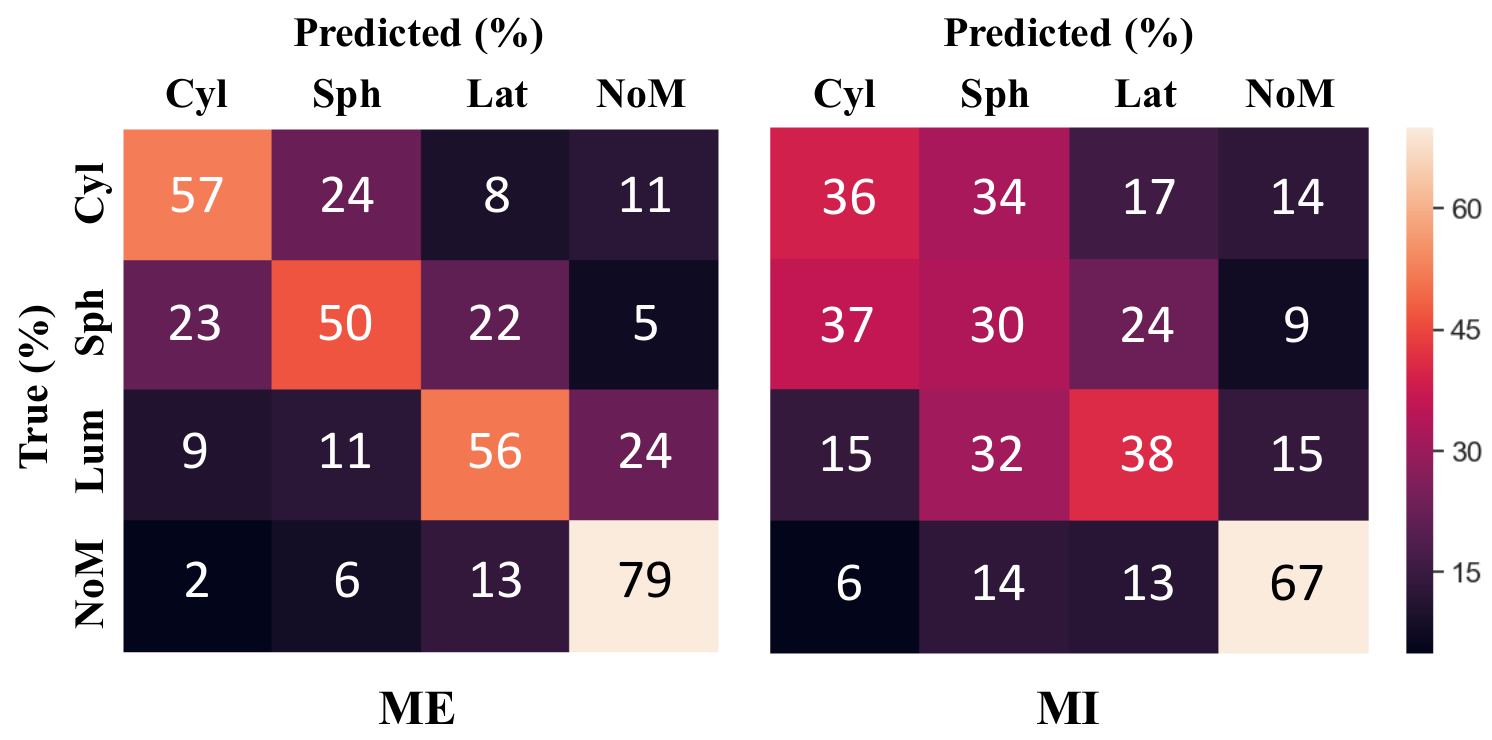}}
\caption{Confusion matrices representing the results of classifying each grasp and no-movement class in ME and MI paradigms.}
\end{figure}

\section {Results and Discussion}
The proposed method, MAP-CNN improved the overall classification performance of the BCI system. We compared MAP-CNN to six other conventional methods, as shown in Fig. 4. These were CSP, CSP$_{MT}$, FBCSP, FBCSP$_{MT}$, EEG-CNN, and EEG-CNN$_{MT}$. We applied multiple time windows (MT) to maximize the performance of each method through the application of various time intervals. In offline decoding, MAP-CNN outperformed the best conventional method with the improvement of 20.9\% and 8.6\% in ME and MI, respectively, for four classes including no-movement.  

\textit{i}) CSP \cite{kim2014decoding, vuckovic2018unimanual}: Features and extracted by CSP on EEG data of time window 0--4 s at [4--40] Hz band. LDA was used to enable the separation of the data into classes. 

\textit{ii}) CSP$_{MT}$ \cite{zhang2018temporally}: Features and extracted by CSP on EEG data of multiple time windows in the [4--40] Hz band. All the features are concatenated for classification.

\textit{iii}) FBCSP \cite{FBCSP, park2017filter, chin2014discriminative}: Optimizes features on EEG data of time window 0--4 s at multiple sub-bands with 4 Hz bandwidth size and 2 Hz step size.

\textit{iv}) FBCSP$_{MT}$ \cite{zhang2018temporally, chin2014discriminative}: Features extracted by FBCSP on EEG data of multiple time windows at multiple sub-bands.  

\textit{v}) EEG-CNN \cite{tabar2016novel, zhang2019novel, EEGNET}: Multi-layers neural networks with a convolution-pooling layer pair and a fully-connected layer at the output. Input images are designed to contain frequency, time, and electrode location information from EEG signals.

\textit{vi}) EEG-CNN$_{MT}$: Multi-layers neural networks on EEG data of multiple time windows.

In the pseudo-online classification, we analyzed the EEG data under the ME and MI paradigms. In Fig. 5, the illustration describes the experimental protocol of pseudo-online. Particular images representing the MAP of each class can be determined by the difference, as shown in Fig. 6. The numbers located below of the images denote modified time intervals after the time-shift. Each image was generated from the data of 4,000 ms length. Thus, the ending should be 3,000 ms if the start was -1,000 ms. The image surrounded by a red border signifies an image of the MAP, estimated by decoding the EEG. Data analysis was performed using nine different time intervals and the predicted image looks more similar to the time-shifted image especially shifted to the past than the original. Hence, we could estimate that the EEG data are related to the past event of muscle movements.

We created the pseudo-online data by concatenating the offline data. We obtained the classification accuracy to each class including no-movement class using confusion matrices in Fig. 7. In summary, the classification results in ME and MI showed a similar tendency. At the same time, we found that the classification accuracy of the no-movement class is higher than other grasping classes. MAP-CNN achieved an averaged accuracy of 60.5($\pm$8.4)\% in ME and 42.7($\pm$6.8)\% in MI on pseudo-online data. The accuracy results showed that the decoding performance of MAP is relatively robust since the result of pseudo-online is very close to the result of the offline. 

\section{Conclusion and Future Works}
This paper showed the feasibility of predicting continual muscle movements with BCI. The proposed method, called MAP-CNN, is based on a CNN architecture and suggests a novel approach to creating MAP images from EEG signals. This approach improved the classification performance in the offline and pseudo-online experiments. Based on the outcome of robust performance confirmed through the offline and pseudo-online experiments, the proposed method will efficiently increase the classification performance, even in BCI systems operated in an online environment. Therefore, we will focus on developing high-performance BCI systems to overcome the limitations identified through this study in future work.  

\section*{Acknowledgment}
The authors thank to D.-H. Lee for help with the database construction and useful discussions of the experiment. 

\bibliographystyle{IEEEbib}
\bibliography{refs}

\end{document}